\documentclass[aps,pre,twocolumn,groupedaddress]{revtex4-1}
\usepackage{graphicx}% Include figure files
\usepackage{dcolumn}% Align table columns on decimal point
\usepackage{bm}% bold math
\usepackage[cp1251]{inputenc}
\usepackage[english]{babel}
\usepackage{epsfig}
\usepackage[cp1251]{inputenc}
\usepackage[mathlines]{lineno}
\usepackage{amsmath,amssymb,graphicx,color}
\usepackage{multibib}

\begin{document}

\title{Superdirective dielectric nanoantennas with effect of light steering}

\author{Alexander~E. Krasnok$^{1}$, Dmitry~S. Filonov$^{1}$, Alexey~P. Slobozhanyuk$^{1}$,
Constantin~R. Simovski$^{2}$, Pavel~A. Belov$^{1}$, Yuri~S. Kivshar$^{1,3}$}
\address{
$^{1}$National Research University of Information Technologies,
Mechanics and Optics, St. Petersburg 197101, Russia\\
$^{2}$Aalto University, School of Electric and Electronic Engineering, Aalto FI76000, Finland \\
$^{3}$Nonlinear Physics Center, Research School of Physics and
Engineering, Australian National University, Canberra ACT 0200,
Australia}

\begin{abstract}
We introduce a novel concept of superdirective antennas based on the
generation of higher-order optically-induced magnetic multipole
modes. All-dielectric nanoantenna can be realized as an optically
small spherical dielectric nanoparticle with a notch excited by a
point source (e.g. a quantum dot) located in the notch. The
superdirectivity effect is not associated with high dissipative
losses. For these dielectric nanoantennas we predict the effect of
the beam steering at the nanoscale characterized by a subwavelength
sensitivity of the beam radiation direction to the source position.
We confirm the predicted effects experimentally through the scaling
to the microwave frequency range.
\end{abstract}

%\pacs{} % PACS, the Physics and Astronomy Classification Scheme.

\maketitle

{\bf Introduction}.
Similar to conventional antennas, a nanoantenna converts a localized electromagnetic
field into freely propagating light, and vice versa~\cite{Lippitz10, Hulst2010,
Novotny_10_NatPhot, Hecht_obz_12}. For optical wireless
circuits on a chip, nanoantennas are required to be both highly
directive and compact~\cite{AluWireless, Hecht_obz_12, Landesa}.
In nanophotonics, directivity has been achieved
for arrayed plasmonic antennas utilizing the
Yagi-Uda design~\cite{Novotny_10_NatPhot, Giessen11,
Hecht_obz_12,KrasnokOE,Chew12}, large dielectric
spheres~\cite{Bonod10}, and metascreen antennas~\cite{Eleftheriades}.
Though individual elements of these arrays are optically small,
the overall size of the radiating systems is larger than
the radiation wavelength $\lambda$.  In addition,
small plasmonic nanoantennas possess weak directivity
close to the directivity of a point dipole~\cite{Jain,
He,Rodriguez}.

Recently, it was suggested theoretically and experimentally to
employ magnetic resonances of high-index dielectric nanoparticles
for enhancing the nanoantenna directivity~\cite{Krasnok_APL_12,
KrasnokOE,BonodOE}. High-permittivity nanoparticles can have nearly
resonant balanced electric and magnetic dipole
responses~\cite{Krasnok_11,KrasnokOE, Evlyukhin, Kuznetsov,
Lukyanchuk13,BonodOE}. This balance of the electric and magnetic
dipoles oscillating with the same phase allows the practical
realization of the Huygens source, an elementary emitting system
with a cardioid pattern~\cite{Balanis,Krasnok_11, Krasnok_APL_12,
KrasnokOE} and with the directivity larger than 3.5. Importantly, a
possibility to excite magnetic resonances leads to the improved
nanoantenna directional properties without a significant increase of
its size.

Superdirectivity has been already discussed for radio-frequency
antennas, and it is defined as directivity of an electrically small
radiating system that significantly exceeds (at least in 3 times)
directivity of an electric dipole~\cite{Balanis, Hansen, Collin}. In
that sense, the Huygens source is not superdirective. In the antenna
literature, superdirectivity is claimed to be achievable only in
antenna arrays by the price of ultimately narrow frequency range and
by employing very precise phase shifters (see, e.g.,
Ref.~\cite{Balanis, Hansen, Collin}). Therefore, superdirective
antennas, though very desirable for many applications such as space
communications and radioastronomy, were never demonstrated and
implemented for practical applications.

\begin{figure}[!b]
\centerline{\includegraphics[width=9cm]{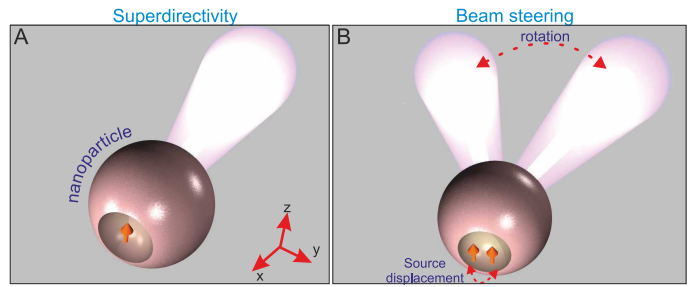}}
\caption{({\bf A}) Geometry of an all-dielectric superdirective
nanoantenna excited by a point-like dipole. ({\bf B}) Concept of the
beam steering effect at the nanoscale.}\label{geometr}
\end{figure}

Superdirectivity was predicted theoretically for an antenna
system~\cite{Eleftheriades} where some phase shifts were required
between radiating elements to achieve complex shapes of the
elements of a radiating system which operates as an antenna
array. In this paper, we employ the properties of subwavelength
particles excited by an inhomogeneous field with higher-order
magnetic multipoles. We consider a subwavelength dielectric
nanoantenna (with the size of 0.4 $\lambda$) with a notch
resonator excited by a point-like emitter located in the notch. The
notch transforms the energy of the generated magneto-dipole Mie
resonance into high-order multipole moments, where the magnetic
multipoles dominate. This system is resonantly scattering i.e.
it is very different from dielectric lenses and usual
dielectric cavities which are large compared to the wavelength.
Another important feature of the notched resonator is huge
sensitivity of the radiation direction to a spatial position of the
emitter. This property leads to a strong beam steering effect and
subwavelength sensitivity of the radiation direction to the source
location. The proposed design of superdirective nanoantennas
may also be useful for collecting single-source radiation, monitoring
quantum objects states, and nanoscale microscopy.

In order to achieve superdirectivity, we should generate
subwavelength spatial oscillations of the radiating
currents~\cite{Balanis, Hansen, Collin}. Then, near fields of the
antenna become strongly inhomogeneous, and the near-field zone
expands farther than that of a point dipole. This results in a
growth of the effective antenna aperture which is associated with
the maximum of directivity $\mbox{D}_{\mbox{max}}=4\pi
\mbox{P}_{\mbox{max}}/\mbox{P}_{\mbox{tot}}$, being defined as
$\mbox{S}=\mbox{D}_{\mbox{max}}\lambda^2/(4\pi)$, where $\lambda$ is
the wavelength of radiation in free space, $\mbox{P}_{\mbox{max}}$
and $\mbox{P}_{\mbox{tot}}$ are the maximum power in the direction
of the radiation pattern and the total radiation power,
respectively. Normalizing the effective aperture $\mbox{S}$ by the
geometric aperture for a spherical antenna $\mbox{S}_0=\pi
\mbox{R}_{\mbox{s}}^2$, we obtain the definition of superdirectivity
in the form~\cite{Hansen, Balanis}:
\begin{eqnarray}
\mbox{S}_{n}=\frac{\mbox{D}_{\mbox{max}}\lambda^2}{4\pi^2\mbox{R}_{\mbox{s}}^2}\gg1
\nonumber
\end{eqnarray}
Practically, the value $\mbox{S}_{n}=4\dots 5$ is sufficient
for superdirectivity of a sphere. In this work, maximum of 6.5
for $\mbox{S}_{n}$ is predicted theoretically for the optical
frequency range, and the value of 5.9 is demonstrated experimentally
for the microwave frequency range.

%%%%%%%%%%%%%%%%%%%%%%%%%%%%%%

\textbf{All--dielectric superdirective optical nanoantennas}.-- Here
we demonstrate a possibility to create a superdirective nanoantenna
without hypothetic metamaterials and plasmonic arrays. We consider a
silicon nanoparticle, taking into account the frequency dispersion
of the dielectric permittivity~\cite{Palik}. The radius of the
silicon sphere is equal in our example to $\mbox{R}_{\mbox{s}}=90$
nm. For a simple sphere under rather homogeneous (e.g. plane-wave)
excitation, only electric and magnetic dipoles can be resonantly
excited while the contribution of higher-order multipoles is
negligible~\cite{KrasnokOE}. Making a notch in the sphere breaks the
symmetry and increases the contribution of higher-order multipoles
into scattering even if the sphere is still excited homogeneously.
Further, placing a nanoemitter (e.g. a quantum dot) inside the
notch, as shown in (Fig.~\ref{geometr}) we create the conditions for
the resonant excitation of multipoles: the field exciting the
resonator is now spatially very non-uniform as well as the field of
a set of multipoles. In principle, the notched particle operating as
a nanoantenna can be performed of different semiconductor materials
and have various shapes -- spherical, ellipsoidal, cubic, conical,
as well as the notch. However, in this work, the particle is a
silicon sphere and the notch has the shape of a hemisphere with a
radius $\mbox{R}_{\mbox{n}}< \mbox{R}_{\mbox{s}}$. The emitter is
modeled as a point-like dipole and it is shown in
(Fig.~\ref{geometr}) by a red arrow.

It is important to mention that our approach is seemingly close to
the idea of Refs.~\cite{Sveta1, Wang} where a small notch on a
surface of a semiconductor microlaser was used to achieve higher
emission directivity by modifying the field distribution inside the
resonator~\cite{Scully}. An important difference between those
earlier studies and our work is that the design discussed earlier is
not optically small and the directive emission is not related to
superdirectivity. In our case, the nanoparticle is much smaller than
the wavelength, and our design allows superdirectivity. For the same
reason our nanoantenna is not dielectric ~\cite{Kim09,Lukyanchuk11}
or Luneburg~\cite{Lipson11,Leonhardt11} lenses. For example,
immersion lenses \cite{Rigneault08,Quake10,Hanson11,Wrachtrup10} are
the smallest from known dielectric lenses, characterized by the
large size 1-2 $\mu m$ in optical frequency range. The functioning
of such lenses is to collecting a radiation by large geometric
aperture $\mbox{S}$, while $\mbox{S}_{n}\simeq1$. Our approach
demonstrates that the subwavelength system, with small geometric
aperture, can have high directing power because of an increase of
the effective aperture. Moreover, there are articles (see. Refs.
\cite{Chew,Klimov}) where the transition rates of atoms inside and
outside big dielectric spheres with low dielectric constant
(approximately 2), were studied.

\begin{figure}[!t]
\centerline{\includegraphics[width=9cm]{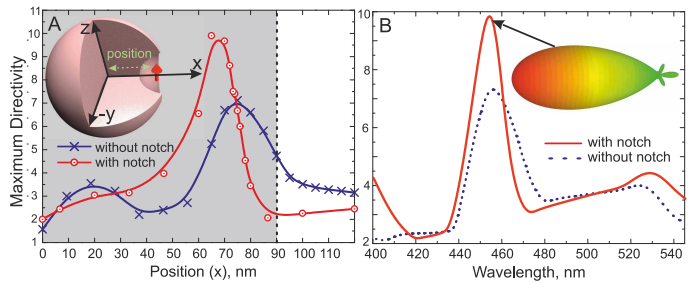}}
\caption{({\bf A}) Maximum of directivity depending on the position
of the emitter ($\lambda=455$ nm) in the case of a sphere with and
without notch. Vertical dashed line marks the particle radius
centered at the coordinate system. ({\bf B}) Directivity dependence
on the radiation wavelength. The inset shows three-dimensional
radiation pattern of the structure ($\mbox{R}_{\mbox{s}}=90$~nm and
$\mbox{R}_{\mbox{n}}=40$~nm).} \label{direct}
\end{figure}

First, we consider a particle without a notch but excited
inhomogeneously by a point emitter. To study the problem
numerically, we employed the simulation software CST Microwave
Studio. Image (Fig.~\ref{direct}A) shows the dependence of the
maximum directivity $\mbox{D}_{\mbox{max}}$ on the position of the
source in the case of a sphere $\mbox{R}_{\mbox{s}}=90$~nm without a
notch, at the wavelength $\lambda=455$~nm (blue curve with crosses).
This dependence has the maximum ($\mbox{D}_{\mbox{max}}=7.1$) when
the emitter is placed inside the particle at the distance 20~nm from
its surface. The analysis shows that in this case the electric field
distribution inside a particle corresponds to the noticeable
excitation of higher-order multipole modes not achievable with the
homogeneous excitation.

Furthermore, the amplitudes of high-order multipoles are
significantly enhanced via making a small notch around the emitter
(see ~\cite{supplementary}). As it is shown in (Fig.~\ref{geometr}),
this geometry transforms it into a resonator for high-order
multipole moments. In this example the center of the notch is on the
nanosphere's surface. The optimal radius of the notch (for maximal
directivity) is equal $\mbox{R}_{\mbox{n}}=40$ nm. In
(Fig.~\ref{direct}A) the extrapolation red curve with circles,
corresponding to simulation results, shows the maximal directivity
versus the location of the emitter at the wavelength 455~nm. The
maximal directivity $\mbox{D}_{\mbox{max}}=10$ is achieved at this
wavelength as one can see from (Fig.~\ref{direct}B) that shows the
directivity versus $\lambda$ with and without a notch. The inset
shows the three-dimensional radiation pattern of the structure at
$\lambda=$455 nm. This pattern has an angular width (at the level of
3 dB) of the main lobe equal to $40^{\circ}$. This value of
directivity corresponds to the normalized effective aperture
$\mbox{S}_{n}=6.5$.

\begin{figure}[!t]
\centerline{\includegraphics[width=9cm]{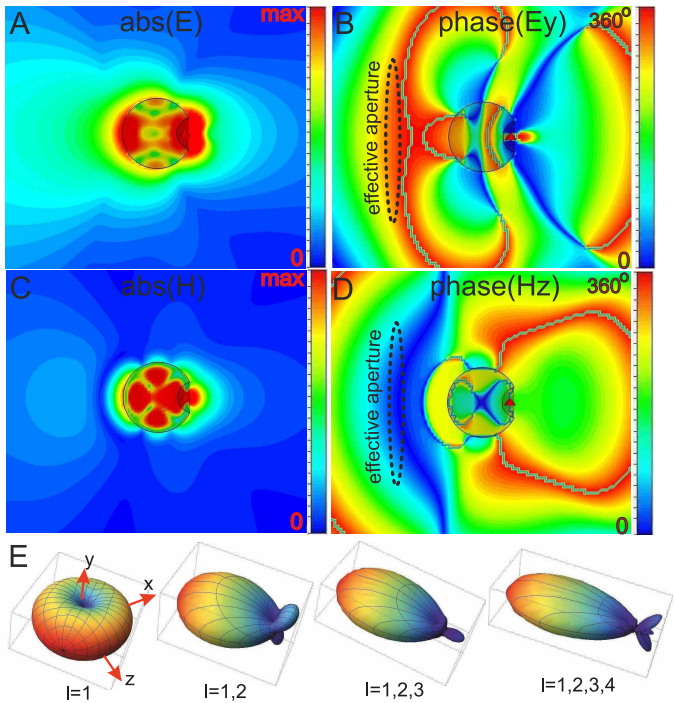}}
\caption{Distribution of ({\bf A}) absolute values and ({\bf B})
phases of the electric field ({\bf C} and {\bf D} for magnetic
field, respectively) of the all-dielectric superdirective
nanoantenna with source in the center of notch, at the wavelength
$\lambda=455$ nm. ({\bf E}) Dependence of the radiation pattern of
all-dielectric superdirective nanoantenna on the number of taken
into account multipoles. Dipole like source located along the $z$
axis.} \label{fieldPatt}
\end{figure}

Figures (Fig.~\ref{fieldPatt}A,B) show the distribution of the
absolute values and phases of the internal electric field and this
field in the vicinity of the nanoantenna. Electric field inside the
particle is strongly inhomogeneous at $\lambda=$455~nm i.e. in the
regime of the maximal directivity (the same holds for the magnetic
field, as shown in Fig.~\ref{fieldPatt}C,D). In this regime, the
internal area where the electric field oscillates with approximately
the same phase turns out to be maximal. This area is located near
the back side of the spherical particle, as can be seen in figure
(Fig.~\ref{fieldPatt}B,D). In other words, the effective near zone
of the nanoantenna in the superdirective regime of is maximal.

Usually, high directivity of plasmonic nanoantennas is
achieved by excitation of higher \textit{electrical} multipole
moments in nanoparticles~\cite{RollyOL, Kall09, Pakizeh12} or
for core-shell resonators consisting of a plasmonic material and a
hypothetic metamaterial which would demonstrate the extreme material
properties in the nanoscale~\cite{Alu}. Although, the values of
directivity achieved for such nanoantennas do not allow superdirectivity,
these studies stress the importance of higher multipoles for
the antenna directivity.

We have performed the transformation of multipole coefficients into
an angular distribution of radiation in accordance to
(\ref{fpattern}) (see ~\cite{supplementary}) by using distribution
of the electric and magnetic fields (Fig.\ref{fieldPatt}A-D) and
determined the relative contribution of each order $l$.
(Fig.~\ref{fieldPatt}E) shows how the directivity grows versus the
spectrum of multipoles with equivalent amplitudes. The right panel
of (Fig.~\ref{fieldPatt}E) nearly corresponds to the inset in
(Fig.~\ref{direct}) that fits to the results shown in
(Fig.~\ref{fieldPatt}E).

Generally, the superdirectivity effect has been accompanied by a
significant increase of the effective near field zone of the antenna
compared to that of a point dipole for which the near zone radius is
equal $\lambda/2\pi$. In the optical frequency range this effect is
especially important, considering the crucial role of the near
fields at the nanoscale.

\begin{figure}[b]
\centerline{\includegraphics[width=9cm]{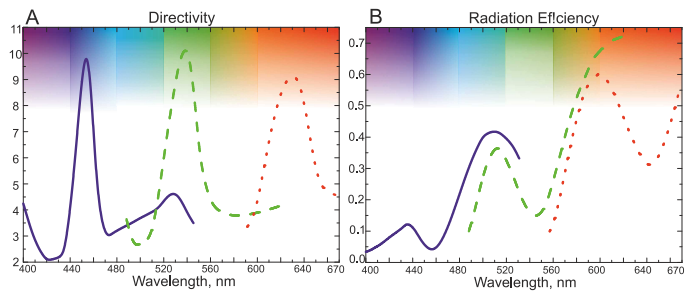}}
\caption{Dependence of directivity ({\bf A}) and radiation
efficiency ({\bf B}) on the size of nanoantenna. Here, the blue
solid lines corresponds to the geometry -- $\mbox{R}_{\mbox{s}}=90$
nm, $\mbox{R}_{\mbox{n}}=40$ nm, the green dashed curves --
$\mbox{R}_{\mbox{s}}=120$ nm, $\mbox{R}_{\mbox{n}}=55$ nm and red
point curves -- $\mbox{R}_{\mbox{s}}=150$ nm,
$\mbox{R}_{\mbox{n}}=65$ nm. Growth of the nanoantenna efficiency
due to the reduction of dissipative losses in silicon with
increasing of wavelength.} \label{directEffDisper1}
\end{figure}

Usually, the superdirectivity regime corresponds to a strong
increase of dissipative losses~\cite{Balanis}. Radiation efficiency
of the nanoantenna is determined by
$\eta_{\mbox{rad}}=\mbox{P}_{\mbox{rad}}/(\mbox{P}_{\mbox{rad}}+\mbox{P}_{\mbox{loss}})$,
where $\mbox{P}_{\mbox{loss}}$ is the power of losses in an
nanoantenna. However, the multipole moments excited in our
nanoantenna are mainly of magnetic type that leads to a strong
increase of the near magnetic field that dominates over the electric
one. Since the dielectric material does not dissipate the magnetic
energy, the effect of superdirectivity does not lead to a so large
increase of losses in our nanoantenna as it would be in the case of
dominating electric multipoles. However, since the electric near
field is nonzero the losses are not negligible. At wavelengths
440-460 nm (blue light) the directivity achieves 10 but the
radiation efficiency is less than 0.1 (see
[Fig.~\ref{directEffDisper1})]. This is because silicon has very
high losses in this range~\cite{Palik}. Peak of directivity is
shifted to longer wavelengths with increasing the size of the
nanoantenna. For the design parameters corresponding to the
operation wavelength 630 nm (red light) the calculated value of
radiation efficiency is as high as 0.5, with nearly same directivity
close to 10. In the infrared range, there are high dielectric
permittivity materials with even lower losses. In principle, the
proposed superdirectivity effect is not achieved by price of
increased losses, and this is an important advantage compared to
known superdirective radio-frequency antenna arrays~\cite{Balanis}
and compared to their possible optical analogues -- arrays of
plasmonic nanoantennas.

%Additional studies have shown that the presence of
%the SiO$_2$ substrate does not lead to a significant reduction of
%the directivity compared to the above results where the nanoantenna is located in free space.
%The absence of the impact of a dielectric substrate is explained by the magnetic nature of the
%nanoantenna operation.

\begin{figure}[t]
\centerline{\includegraphics[width=9cm]{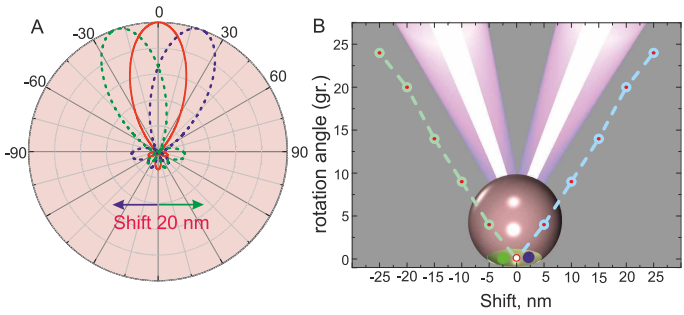}} \caption{
The rotation effect of the main beam radiation pattern, with
subwavelength displacement of emitter inside the notch. ({\bf A})
The radiation patterns of the antenna with the source in center
(solid line) and the rotation of the beam radiation pattern for the
20 nm left/right offset (dashed lines). ({\bf B}) Dependence of the
the rotation angle on the source offset.} \label{shiftEffect}
\end{figure}

\textbf{Steering of light at the nanoscale}.-- Here we examine the
response of the nanoantenna to subwavelength displacements of the
emitter. Displacement in the plane perpendicular to the axial
symmetry of antenna (i.e. along the $y$ axis) leads to rotation of
the beam without damaging the superdirectivity. Image
(Fig.\ref{shiftEffect}A) shows the radiation patterns of the antenna
with the source in center (solid line) and the rotation of the beam
for the 20 nm left/right offset (dashed lines). Shifting of the
source right side leads to rotation of pattern to the left, and vice
versa. The angle of the beam rotation is equal to 20 degrees, that
is essential and available to experimental observations. The result
depends on the geometry of the notch. For a hemispherical notch, the
dependence of the rotation angle on the displacement is presented in
Fig.\ref{shiftEffect}B.

Instead of the movement of a single quantum dot one can have in mind
the emission of two or more quantum dots located near the edges of
the notch. In this case, the dynamics of their spontaneous decay
will be well displayed in the angular distribution of the radiation.
This can be useful for quantum information processing and for
biomedical applications.

Beam steering effect described above is similar to the effect of
beam rotation in hyperlens \cite{Zhang07, Narimanov06, LiuReview12},
where the displacement of a point-like source leads to a change of
the angular distribution of the radiation power. However, in our
case, the nanoantenna has subwavelength dimensions and therefore it
can be neither classified as a hyperlens nor as a micro-spherical
dielectric nanoscope \cite{Kim09,Lukyanchuk11}, moreover it is not
an analogue of solid immersion micro-lenses
\cite{Rigneault08,Quake10,Hanson11,Wrachtrup10}, which are
characterized by the size 1-5 $\mu m$ in the same frequency range.
These lens has a subwavelength resolving power due to the large
geometric aperture but the value of normalized effective aperture is
$\mbox{S}_{n}\simeq1$. Our study demonstrates that the
sub-wavelength system, with \emph{small compared to the wavelength}
geometric aperture can have both high directing and resolving power
\emph{because of a strong increase of the effective aperture
compared to the geometrical one}.

\begin{figure}[t]
\centerline{\includegraphics[width=9cm]{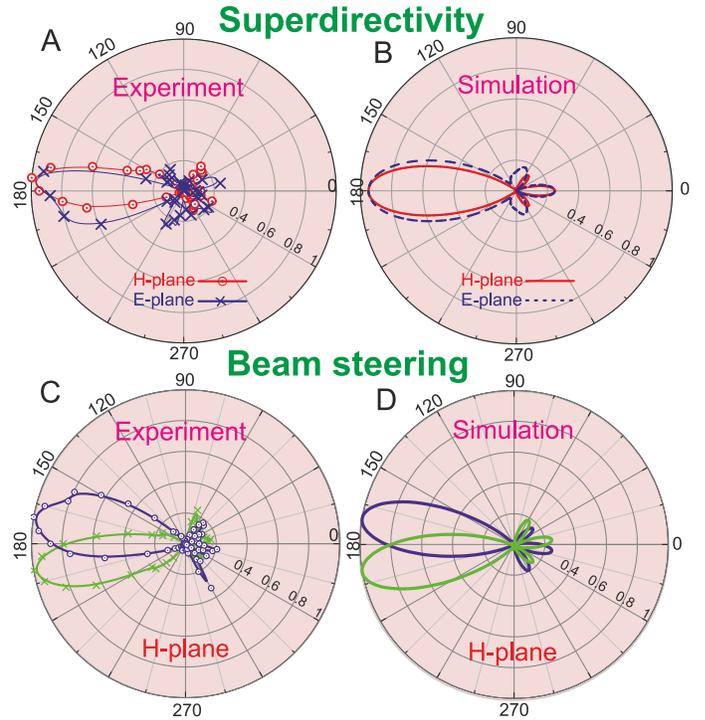}}
\caption{Experimental ({\bf A}) and numerical ({\bf B}) radiation
patterns of the antenna in both $E$- and $H$-planes at the frequency
16.8 GHz. The crosses and circles correspond to the experimental
data. Experimental ({\bf C}) and numerical ({\bf D}) demonstration
of beam steering effect, displacement of dipole is equal 0.5 mm.}
\label{experiment}
\end{figure}

\textbf{Experimental verification}.--We have confirmed both
predicted effects studying the similar problem for the microwave
range. To do this, we have scaled up the nanoantenna as above to low
frequencies. Instead of Si we employ MgO-TiO$_{2}$ ceramic
~\cite{Krasnok_APL_12} characterized at microwaves by a
dispersion-less dielectric constant 16 and dielectric loss factor of
1.12$\cdot$10$^{-4}$. The results of the experimental investigations
(details of which are described in the supplementary materials
\cite{supplementary}) and numerical simulations of the pattern in
both $E$- and $H$-planes are summarized in
(Figs.~\ref{experiment}A,B). Radiation patterns in both planes are
narrow beams with a lobe angle about 35$^{\circ}$. Experimentally
obtained coefficients of the directivity in both $E$- and $H$-planes
are equal to 5.9 and 8.4, respectively (theoretical predictions for
them were equal, respectively, 6.8 and 8.1). Our experimental data
are in a good agreement with the numerical results except a small
difference for the E plane, that can be explained by the imperfect
symmetry of the emitter. Note, that the observed directivity is
close to that of an all-dielectric Yagi-Uda antenna with overall
size $2\lambda$~\cite{Krasnok_APL_12}. The total size of our
experimental antenna is closed to $\lambda/2.5$. Thus, our
experiment clearly demonstrates the superdirective effect.

Experimental and numerical demonstration of the beam steering effect
are presented in (Figs.~\ref{experiment}C,D). For the chosen
geometry of antenna, displacement of source by 0.5 mm leads to a
rotation of the beam about 10$^{\circ}$. Note that the ratio of
$\lambda=18.7$ mm to value of the source displacement 0.5 mm is
equal to 37. This proves that the beam steering effect observed at
subwavelength displacement of source.

\textbf{Conclusions}.~ We have suggested a novel approach to achieve
superdirectivity of antennas through the excitation of
higher-order magnetic multipoles in an optically small dielectric
nanoparticle with a noth and a point emitter located inside the
notch. For the visible frequency range, we have studied this effect
theoretically and demonstrated that a nanoemitter placed in the notch
generates efficiently higher-order magnetic miltipole modes
responsible for a very high directivity of the nanoantenna
not achievable by any other method. We have also suggested
an efficient steering effect for an offset of the subwavelength source.
We have demonstrated experimentally both superdirectivity and
giant beam steering for the microwave frequency range. Combination
of superdirectivity with the beam steering and the fact that both the effects
can be observed for optical and microwave ranges makes our results very promising
for numerous applications in radio physics and nanophotonics.

\textbf{Acknowledgements.} The authors thank B. Lukyanchuk, S.~A.
Tretyakov and V.~V. Klimov for useful discussions and suggestions.
The authors thank also E.~A. Nenasheva and P.~V. Kapitanova for a
technical help. This work was supported by the Ministry of Education
and Science of the Russian Federation (projects 11.G34.31.0020,
11.519.11.2037, 14.B37.21.0303, 14.132.21.1678, 01201259765),
Dynasty Foundation (Russia), and the Australian Research Council.

%\bibliography{liter}

\clearpage
\section{Supplementary Material}\label{Supplementary}
\renewcommand\figurename{S.}
\setcounter{figure}{0}
\setcounter{equation}{0}

\textbf{Multipole expansion for superdirectivity}. ~In this section,
we demonstrate how to find multipole modes excited in the
all-dielectric superdirective nanoantenna which are responsible for
its enhanced directivity. We expand the exactly simulated internal
field, producing the polarization currents in the nanoparticle, into
multipole moments following to \cite{Jackson1}. The expansion is a
series of spherical harmonics with the coefficients $a_{E}(l,m)$ and
$a_{M}(l,m)$, which characterize the electrical and magnetic
multipole moments \cite{Jackson1}:

\begin{eqnarray}
&&a_{E}(l,m)=\frac{4\pi
k^2}{i\sqrt{l(l+1)}}\int{Y_{lm}^{*}\left[\rho\frac{\partial}{\partial
r}[rj_{l}(kr)]+\frac{ik}{c}(\mathbf{r}\cdot\mathbf{j})j_{l}(kr)\right]}d^{3}x \nonumber \\
&&a_{M}(l,m)=\frac{4\pi
k^2}{i\sqrt{l(l+1)}}\int{Y_{lm}^{*}\mbox{div}\left(\frac{\mathbf{r}\times
\mathbf{j}}{c}\right)j_{l}(kr)}d^{3}x \label{moments}
\end{eqnarray}

where $\rho=1/(4\pi)\mbox{div}(\mathbf{\mbox{E}})$ and
$\mathbf{j}=c/(4\pi)(\mbox{rot}(\mathbf{\mbox{H}})+ik\mathbf{\mbox{E}})$
are densities of the induced electrical charges and polarization
currents that can be easily expressed through the internal electric
$\mathbf{\mbox{E}}$ and magnetic $\mathbf{\mbox{H}}$ fields of the
sphere, $Y_{lm}$ - spherical harmonics of the orders $(l>0,0\ge
|m|\le l)$, $k=2\pi/\lambda$, $j_{l}(kr)$ - spherical Bessel
function order $l$ and $c$ is the speed of light. Coefficients
$a_{E}(l,m)$ and $a_{M}(l,m)$ determine the electric and magnetic
mutipole moments, namely dipole at $l=1$, quadrupole at $l=2$,
octupole at $l=3$ etc..

\begin{figure}[h]
\centerline{\includegraphics[width=9cm]{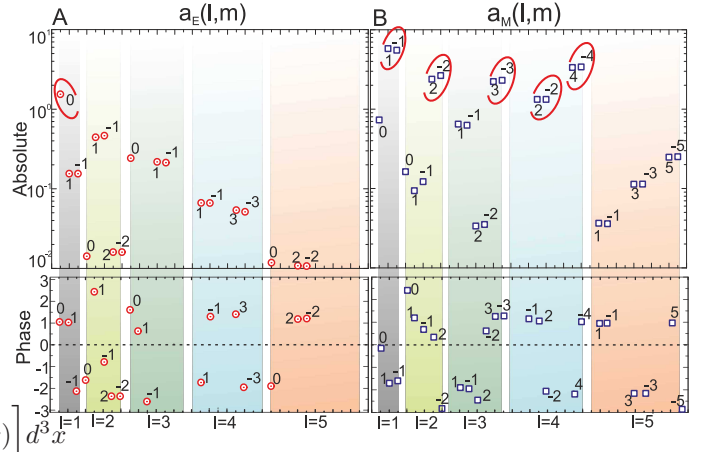}}
\caption{Absolute values and phases of ({\bf A}) electric and ({\bf
B}) magnetic multipole moments that provide the main contribution to
the radiation of all-dielectric superdirective optical nanoantenna
at the wavelength 455~nm. Multipole coefficients providing the
largest contribution to the antenna direction are highlighted by red
circles.} \label{allmoments1}
\end{figure}

The multipole coefficients determine not only the mode structure of
the internal field but also the angular distribution of the
radiation. In particular, in the far field zone electric and
magnetic fields of multipole order $l$ depend on the distance $r$ as
$\sim(-1)^{i+1}\frac{\exp(ikr)}{kr}$ and expression for the angular
distribution of the radiation power can be written as follows:
\begin{widetext}
\begin{eqnarray}
&&\mbox{d}P(\theta,\varphi)=\frac{c}{8\pi
k^2}\left|\sum_{l,m}(-i)^{l+1}[a_{E}(l,m)\mathbf{X}_{lm}\times\mathbf{n}+a_{M}(l,m)\mathbf{X}_{lm}]\right|^2\mbox{d}\Omega,\nonumber\\
&&\mathbf{X}_{lm}(\theta,\varphi)=\frac{1}{\sqrt{l(l+1)}}
\begin{bmatrix}
1/2(\sqrt{(l-m)(l+m+1)}Y_{l,m+1}+\sqrt{(l+m)(l-m+1)}Y_{l,m-1})\\
1/(2i)(\sqrt{(l-m)(l+m+1)}Y_{l,m+1}-\sqrt{(l+m)(l-m+1)}Y_{l,m-1})\\
mY_{l,m}
\end{bmatrix}.
\label{fpattern}
\end{eqnarray}
\end{widetext}

Here
$\mbox{d}\Omega=\mbox{sin}(\theta)\mbox{d}\theta\mbox{d}\varphi$ is
the solid angle element in spherical coordinates and $\mathbf{n}$ -
unit vector of the observation point. All coefficients $a_E(l,m)$
and $a_M(l,m)$ give the same contribution to the radiation, if they
have the same values. Since higher-order multipoles for optically
small systems have usually negligibly small amplitudes compared to
$a_E(1,m)$ and $a_M(1,m)$, they are, as a rule, not considered.

The amplitudes of multipole moments for electric and magnetic fields
distribution (Fig.\ref{fieldPatt}A-D) are shown in figure
(S.~\ref{allmoments1}), where we observe strong excitation of
$a_{E}(1,0)$, $a_{M}(1,1)$, $a_{M}(1,-1)$, $a_{M}(2,2)$,
$a_{M}(2,-2)$, $a_{M}(3,3)$, $a_{M}(3,-3)$, $a_{M}(4,2)$,
$a_{M}(4,-2)$, $a_{M}(4,4)$ and $a_{M}(4,-4)$. These multipole
moments determine the angular pattern of the antenna. All other ones
give a negligible contribution. Absolute values of all magnetic
moments are larger than those of the electric moments in the
corresponding multipole orders, and the effective spectrum of
magnetic multipoles is also broader than that of the electric
moments. Thus, the operation of the antenna is mainly determined by
the magnetic multipole response. Absolute values of multipole
coefficients $a_{M}(l,\pm |m|)$ of the same order $l$ are
practically equivalent, however, the phase of some coefficients are
different. Therefore, the modes with $+|m|$ and $-|m|$ form a strong
anisotropy of the forward--backward directions that results in the
unidirectional radiation.

\begin{figure}[!t]
\centerline{\includegraphics[width=9cm]{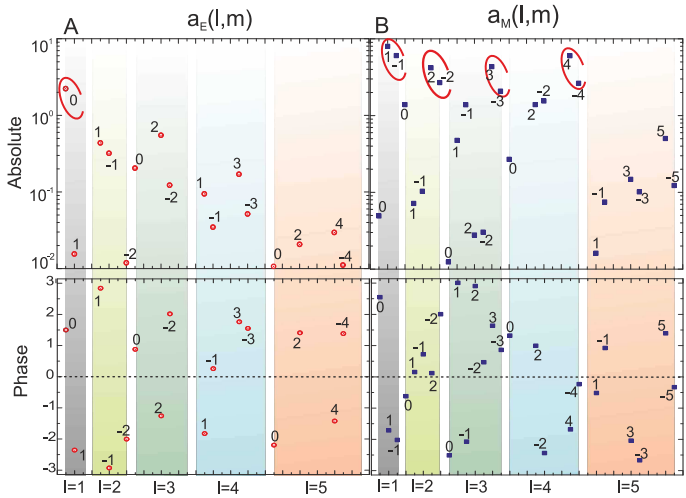}}
\caption{Absolute values and phases of ({\bf A}) electric and ({\bf
B}) magnetic multipole moments that provide the main contribution to
the radiation of all-dielectric superdirective optical nanoantenna
in case of asymmetrical location of source at the wavelength 455~nm.
Coefficients that give the largest contribution to the antenna
directivity are highlighted by red circles.} \label{shiftmagnet}
\end{figure}

\textbf{Multipole expansion for beam steering}. ~To interpret beam
steering effect, consider the result of field expansion to magnetic
multipoles, as shown in (S.\ref{shiftmagnet}). In the case of
asymmetrical location (the 20 nm left offset) of the source in the
notch absolute values of $a_{M}(l,\pm |m|)$ are different. This
means that the mode $a_{M}(l,+|m|)$ is excited more strongly than
$a_{M}(l,-|m|)$, or vice versa, that depends on direction of
displacement. The effect of superdirectivity remains at offset of
source until to the edge of notch. Small displacements of the source
along $x$ and $z$ do not lead to the rotation of the pattern.

\textbf{Details and technique of experiment}

\begin{figure}[!b]
\centerline{\includegraphics[width=9cm]{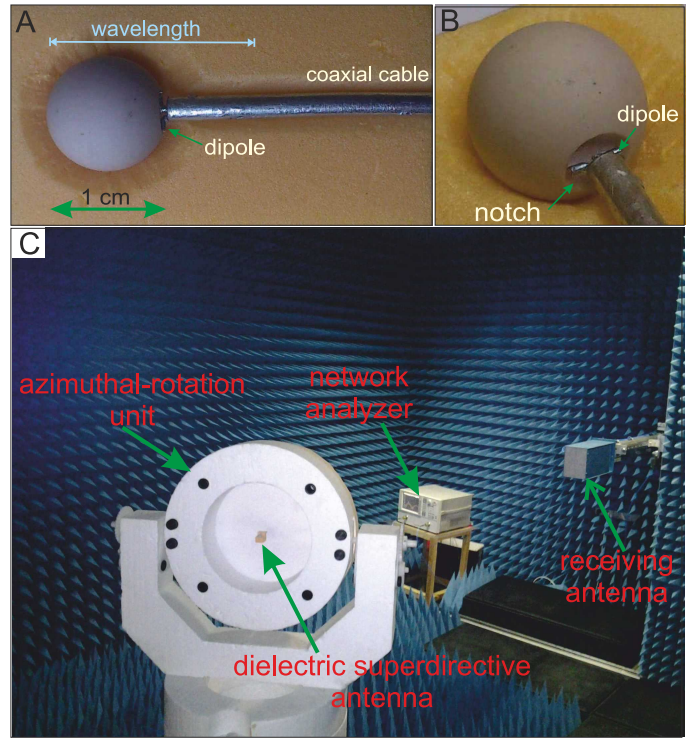}}
\caption{Photographs of ({\bf A}) top view and ({\bf B}) perspective
view of a notched all-dielectric microwave antenna. Image of ({\bf
C}) the experimental setup for measuring of power patterns.}
\label{Geomexper}
\end{figure}

This topic provides some details and technique of the experiments.
For confirmed superdirectivity and beam steering effects, we have
scaled up the nanoantenna as above to low frequencies. Instead of Si
we employ MgO-TiO$_{2}$ ceramic ~\cite{KrasnokAPL} characterized at
microwaves by a dispersion-less dielectric constant 16 and
dielectric loss factor of 1.12$\cdot$10$^{-4}$. We have used the
sphere of radius $\mbox{R}_{\mbox{s}}=5$~mm and applied a small wire
dipole~\cite{Bal} excited by a coaxial cable [see
(S.~\ref{Geomexper}A,B)]. The common-mode radiation (i.e. that of
the cable) was prevented with a mini-balun located inside the
sphere. The size of the hemispherical notch is approximately equal
to $\mbox{R}_{\mbox{n}}=2$ mm. Styrofoam material with the
dielectric permittivity close to 1 is used to fix the antenna in the
azimuthal-rotation unit, as shown in (S.~\ref{Geomexper}C). First,
we have performed a full-wave simulations of the whole structure
antenna (with the CST Microwave Studio), and observed the
superdirectivity at the frequency 16.8 GHz. Next, we have studied
experimentally the radiation pattern of this antenna in the anechoic
chamber. Radiating power was measured by ultra-wideband antenna TMA
(1.0-18.0 HF) which was located in the far field radiation zone. The
results of the experimental investigations and numerical simulations
of the pattern in both $E$- and $H$-planes are summarized in
(Figs.~\ref{experiment}A,B) [see main part]. Radiation patterns in
both planes are narrow beams with a lobe angle about 35$^{\circ}$.
Experimentally obtained coefficients of the directivity in both $E$-
and $H$-planes are equal to 5.9 and 8.4, respectively (theoretical
predictions for them were equal, respectively, 6.8 and 8.1). Our
experimental data are in a good agreement with the numerical results
except a small difference for the E plane, that can be explained by
the imperfect symmetry of the wire dipole.

%\begin{figure}[!t]
%\centerline{\includegraphics[width=9cm]{S11exper.eps}}
%\caption{Level of the return losses of superdirective dielectric
%antenna. Blue area shows the operating frequency range.}
%\label{S11exper}
%\end{figure}

Experimental and numerical demonstration of the beam steering effect
at microwave are presented in (Figs.~\ref{experiment}C,D) [see main
part]. For the chosen geometry of antenna, displacement of source by
0.5 mm leads to a rotation of the beam about 10$^{\circ}$. Note that
the ratio of $\lambda=18.7$ mm to value of the source displacement
0.5 mm is equal to 37. This proves that the beam steering effect
observed at subwavelength displacement of source.

%Finally, we consider the question of dielectric superdirective
%antenna matching with coaxial cable. Despite that length of the wire
%dipole close to $\lambda/10$, dielectric superdirective antenna is
%well agreement with coaxial cable in the operating frequency range.
%The figure (S.\ref{S11exper}) shows the level of return loss for
%this case. This well agreement is explained by the strong coupling
%of the wire dipole with the excited modes of notched dielectric
%particle and is not related to the dissipative losses in the
%superdirectivity regime. For this reason, we have not used
%additional matching devices (e.g. "balun").

\end{document}